\documentclass{article}
\usepackage{spconf,amsmath,amssymb,graphicx,hyperref}
\usepackage{booktabs,array,multirow,xcolor,colortbl}
\usepackage{tikz}
\usepackage{pgfplots}
\usepackage{arydshln}
\usepackage{enumitem}

\definecolor{orcidlogocol}{HTML}{A6CE39}
\newif\ifhasorcidlink
\IfFileExists{orcidlink.sty}{\usepackage{orcidlink}\hasorcidlinktrue}{\hasorcidlinkfalse}
\newcommand{\orcidicon}[1]{%
  \ifhasorcidlink
    \orcidlink{#1}%
  \else
    \href{https://orcid.org/#1}{%
      \tikz[baseline=-0.55ex]{\node[circle,fill=orcidlogocol,inner sep=0.35pt,text=white,font=\sffamily\bfseries\tiny]{iD};}%
    }%
  \fi
}
\newcommand{\orcid}[1]{%
  \begingroup
  \hypersetup{pdfborder={0 0 0}}%
  \,\textnormal{\orcidicon{#1}}%
  \endgroup
}

\pgfplotsset{compat=1.14}
\usetikzlibrary{arrows.meta,positioning,fit,backgrounds,decorations.pathreplacing}

\definecolor{cMerge}{HTML}{1B6CA8}
\definecolor{cBase}{HTML}{B0562B}
\definecolor{cAccent}{HTML}{4E7A51}
\definecolor{cGrey}{HTML}{6B6B6B}
\definecolor{hlrow}{HTML}{EAF2F8}
\definecolor{cOther}{HTML}{A8D5A2}
\definecolor{cMergeRed}{HTML}{E78AA0}

\newcommand{\best}[1]{\textbf{#1}}
\newcommand{\dW}{\Delta\mathbf{W}}
\newcommand{\obs}[1]{\smallskip\noindent\textbf{#1}\hspace{2pt}}

\title{Dual Task Vector Merging: Leveraging Emotion and ASR Knowledge\\ for Low-Resource Alzheimer's Disease Detection}
\name{\shortstack{
Yuan-Chao Li$^1$\orcid{0000-0003-4266-2005},
Yin-Long Liu$^2$\orcid{0009-0004-0380-0836},
Liu He$^2$\orcid{0000-0002-6128-4292},
Sizhe Liu$^2$\orcid{0009-0003-6405-6492},
Ji-Xuan Deng$^2$\orcid{0009-0007-3910-3911}, \\
Yin Zhang$^2$\orcid{0009-0008-2933-0544},
Zhen-Hua Ling$^2$\orcid{0000-0001-7853-5273},
Jiahong Yuan$^2$\orcid{0009-0008-2162-7167}
}}
\address{$^1$University of Cambridge, UK \quad $^2$University of Science and Technology of China, China}

\begin{document}
\ninept
\maketitle

\begin{abstract}
Alzheimer's Disease (AD) detection from speech benefits from both paralinguistic and linguistic cues, yet model training is frequently constrained by limited data. Fine-tuning a large pretrained speech encoder can easily overfit, while a frozen probe leaves the encoder's task-relevant knowledge underutilized. Therefore, we propose \emph{Dual Task Vector Merging}, a parameter-efficient method that adapts a frozen wav2vec 2.0 backbone using task knowledge learned outside the AD task. Specifically, we inject two task vectors, derived as weight deltas from emotion- and ASR-fine-tuned wav2vec 2.0 models, into each transformer layer. During AD training, only layer-wise merging coefficients and a linear classification head are optimized, updating less than $0.1\%$ of the parameters required for full fine-tuning. Speaker-independent experiments on ADReSS and ADReSSo show that dual merging improves over frozen probes, LoRA, and full fine-tuning, in the main binary settings, and gives the best results when combined with LoRA. Evaluation on PROCESS further shows that our advantages transfer to three-class HC/MCI/dementia classification when paired with LoRA. Experiments under reduced training data (20\%, 40\%, and 80\%) confirm reliable gains in low-resource settings\footnote{Code available upon acceptance.}.
\end{abstract}


\section{Introduction}
\label{sec:intro}

Alzheimer's Disease (AD) causes progressive cognitive decline and is often accompanied by changes in speech production, including lexical retrieval difficulty, disfluency, altered timing, and degraded articulation. Compared with neuroimaging and other clinical procedures, speech offers a low-cost and non-invasive route to early screening~\cite{delafuente2020review}, and has therefore become a central signal in recent AD detection challenges such as ADReSS and ADReSSo~\cite{luz2020adress,luz2021adresso}. Modern systems commonly rely on Self-Supervised Learning (SSL) encoders such as wav2vec 2.0, either by freezing the encoder and training a light probe, or by fine-tuning the full backbone~\cite{baevski2020wav2vec2,hsu2021robust}.

Despite strong progress, acoustic-only AD detection remains a low-resource problem. Benchmark corpora typically contain around one hundred labeled speakers, and speaker-independent evaluation leaves even fewer subjects for each training fold. This creates a sharp trade-off. A frozen probe is robust but can only read whatever AD-relevant information is already available in the pretrained representation. Fine-tuning can reshape the representation, but exposes millions of parameters to a small clinical dataset and is vulnerable to overfitting. The problem is not only computational cost: in small AD folds, additional capacity can reduce rather than improve generalization.

We ask whether the encoder can be adapted without learning new backbone weights from the AD data. Our starting point is that AD speech is not an isolated acoustic phenomenon. Cognitive decline affects \emph{how} speech is produced, through prosody, affect, and voice quality, and \emph{what} is produced, through fluency and articulation. These dimensions are already emphasized by speech emotion recognition and Automatic Speech Recognition (ASR) models. Task arithmetic makes such knowledge reusable: the difference between a fine-tuned checkpoint and its base model, $\dW=\theta_{\mathrm{ft}}-\theta_{\mathrm{base}}$, can act as a transferable task vector~\cite{ilharco2023task}. Model-merging work further shows that such vectors can be combined in weight space without adding inference-time branches~\cite{wortsman2022soups,matena2022fisher}, while recent adaptive merging methods reduce interference among task directions~\cite{yadav2023ties,yang2024adamerging}.

In this paper, we introduce \textbf{Dual Task Vector Merging} for AD detection. We keep a wav2vec 2.0-Large-Robust encoder frozen and inject two precomputed task vectors, one from emotion and one from ASR, into every transformer layer. Each layer learns only two scalar coefficients, controlling the strength of the emotion and ASR directions, and the downstream AD classifier is a single linear head over a middle-layer representation. The method therefore adapts the effective weights seen at inference while training only approximately 2.1K parameters. Unless otherwise stated, all performance numbers are reported as percentages with the percent sign omitted. Our contributions are:
\begin{itemize}
\itemsep2pt
\item We formulate AD detection as low-dimensional task-vector adaptation of a frozen SSL speech encoder, using complementary emotion and ASR vectors rather than backbone fine-tuning.
\item On ADReSS and ADReSSo, dual merging improves over the frozen probe, LoRA, and full fine-tuning, and uses less than 0.1\% as many trainable parameters as full fine-tuning.
\item We further evaluate the method under reduced training data and on the harder PROCESS three-class benchmark, showing that task-vector priors transfer beyond the standard binary AD/HC setting. 
\end{itemize}

\section{Related Work}
\label{sec:related}

\subsection{Speech-Based Alzheimer's Disease Detection}
Speech-based AD detection has been studied through both linguistic and acoustic cues, including lexical richness, pauses, fluency, prosody, and learned speech representations~\cite{delafuente2020review,cummins2020comparison}. Challenge corpora such as ADReSS and ADReSSo have made speaker-independent benchmark evaluation common~\cite{luz2020adress,luz2021adresso}. Speech-only systems have further explored acoustic feature sets, pretrained speech embeddings, ASR-derived representations, and wav2vec/BERT pipelines~\cite{gauder2021speech_embeddings,zhu2021wavbert,pappagari2021lowresource}. Recent work also studies low-resource cognitive-state classification with multi-view pseudo-labeling~\cite{li2025semicog} and shows that ASR errors themselves may contain useful diagnostic cues for AD detection~\cite{liu2026asrerrors}. Acoustic-only systems are appealing because they do not rely on manual transcripts and can operate directly on recorded speech. This setting also exposes the fragility of small clinical benchmarks. Prior work has shown that apparently strong AD classifiers can exploit shortcut cues such as silence and recording artifacts (i.e., clever hans effect)~\cite{liu2024cleverhans}. This motivates capacity-controlled methods that are explicit about which speech knowledge is being transferred.

\subsection{Task Vectors and Model Merging}
Operating directly in the weight space, model merging provides a parameter-efficient framework to combine distinct task capabilities without inference-time overhead~\cite{wortsman2022soups,matena2022fisher}. By manipulating task vectors derived from fine-tuned checkpoints~\cite{ilharco2023task}, recent approaches move beyond static arithmetic toward interference-aware or adaptive merging coefficients for improved performance~\cite{yadav2023ties,yang2024adamerging}. Task-vector and model-merging ideas have also been explored in speech processing, including ASR adaptation~\cite{ramesh2024task,lin2026speechft} and speech emotion recognition~\cite{adaltm2026}. Prior speech studies typically rely on a single task vector or operate within a single task domain. In contrast, we transfer two domain-distinct speech-task directions, paralinguistic and linguistic, to a third clinical task where direct supervision is scarce.

\subsection{Parameter-Efficient Adaptation}
Parameter-efficient methods such as LoRA learn small adapter modules while keeping the original backbone frozen~\cite{hu2022lora}. They are effective, but still learn new parameters from the target data. Task-vector merging differs in that the directions are learned elsewhere. AD training only chooses how much of each existing speech-task direction to use. This makes merging attractive for clinical datasets, where useful priors and a small hypothesis space are both desirable. The two approaches are not mutually exclusive. We evaluate LoRA alone and LoRA on top of dual vector merging to separate the contribution of reused task knowledge from additional adapter capacity.

\section{Method}
\label{sec:method}

\subsection{Task-Vector Construction}
Let $\theta_{\mathrm{base}}$ denote the pretrained SSL speech encoder and $\theta_t^{\mathrm{ft}}$ denote a checkpoint fine-tuned on task $t$. We define its task vector as
\begin{equation}
    \dW_t = \theta_t^{\mathrm{ft}}-\theta_{\mathrm{base}}.
\end{equation}
We construct two vectors on the same wav2vec 2.0~\cite{hsu2021robust}\footnote{https://huggingface.co/facebook/wav2vec2-large-robust} backbone: $\dW_{\mathrm{emo}}$ from the emotion-msp speech-emotion checkpoint\footnote{https://huggingface.co/audeering/wav2vec2-large-robust-12-ft-emotion-msp-dim} and $\dW_{\mathrm{asr}}$ from the libri-960h ASR checkpoint\footnote{https://huggingface.co/facebook/wav2vec2-large-robust-ft-libri-960h}. Only matching transformer weights are used. The task vectors are stored as fixed tensors and receive no gradient during AD training.

The two source tasks are chosen to cover complementary aspects of pathological speech. The emotion vector emphasizes paralinguistic variation such as prosody and voice quality, while the ASR vector emphasizes phonetic and lexical recoverability. Dual merging lets the AD classifier reuse both directions while leaving the AD backbone unchanged.

\begin{figure}[t]
\centering
\includegraphics[width=\linewidth]{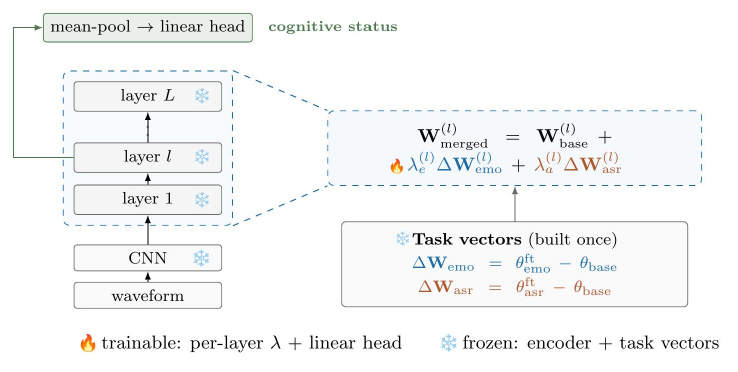}
\caption{Overview of Dual Task Vector Merging. Emo and ASR task vectors are computed once by subtracting the base wav2vec 2.0 weights from task-fine-tuned checkpoints. During AD training, the backbone and task vectors are frozen. Only the layer-wise coefficients $\lambda_e^{(l)}$, $\lambda_a^{(l)}$ and the linear head are updated.}
\label{fig:method}
\vspace{-5pt}
\end{figure}

\subsection{Layer-Wise Dual Merging}
Fig.~\ref{fig:method} illustrates the proposed merging pipeline.
For each transformer layer $l$, we merge the base weight with the two task vectors:
\begin{equation}
\mathbf{W}_{\mathrm{merged}}^{(l)}
=
\mathbf{W}_{\mathrm{base}}^{(l)}
+ \lambda_e^{(l)}\dW_{\mathrm{emo}}^{(l)}
+ \lambda_a^{(l)}\dW_{\mathrm{asr}}^{(l)} .
\label{eq:merge}
\end{equation}
The same pair of coefficients is shared by the layer's query, key, value, output, and feed-forward projections. The coefficients are initialized to $0.5$ and are unconstrained real-valued scalars. In practice, the merged weight is assembled on the fly inside each linear projection:
\begin{equation}
    \mathbf{y}=\mathbf{W}_{\mathrm{merged}}^{(l)}\mathbf{x}+\mathbf{b}.
\end{equation}
Thus the effective encoder is changed, but the underlying pretrained weights and the two task vectors remain frozen. At inference, the weighted deltas can be folded into the layer weights, so the method introduces no additional network branch.

\subsection{AD Classifier and Capacity}
We extract hidden states from a single middle transformer layer (layer 12), apply masked temporal mean pooling over valid frames, and train a linear classifier for cognitive status. For the 24-layer wav2vec 2.0-Large-Robust encoder, dual merging introduces $2\times24=48$ scalar coefficients. Together with the linear head, the full trainable budget is approximately 2.1K parameters, compared with 311M trainable parameters for full fine-tuning. This restricts adaptation to a low-dimensional affine subspace spanned by two speech-task directions per layer.

\section{Experiments}
\label{sec:exp}

\subsection{Experimental Setup}
\noindent\textbf{Datasets.}
We evaluate on three AD speech benchmarks. ADReSS contains age- and gender-balanced Cookie Theft recordings for binary AD/Healthy-Control (HC) classification~\cite{luz2020adress}. ADReSSo is a speech-only follow-up benchmark with full recordings. We keep participant speech and remove interviewer segments~\cite{luz2021adresso}. PROCESS provides a harder three-class setting with HC, Mild Cognitive Impairment (MCI), and dementia labels, and is used as a generalization stress test~\cite{tao2024process}.

\noindent\textbf{Protocol.}
All experiments use speaker-independent 5-fold cross-validation. Fold assignment is performed at the speaker level before chunking, so every chunk from the same speaker is assigned to the same fold. For each held-out test fold, one non-test fold is used as an inner validation fold and the remaining folds are used for training. Models are trained with chunk-level labels, but evaluation is performed at the speaker level. Chunk posterior probabilities are averaged over all chunks from a speaker, and the final prediction is obtained from the averaged probability vector. ADReSS and ADReSSo main results are averaged over five random seeds and five folds. PROCESS, ablations, and low-resource experiments use three seeds and five folds. Each comparison is paired by dataset split, fold, and seed. We report mean$\pm$standard deviation (std) over the paired seed-fold runs as descriptive evidence, not as a formal significance test.

For low-resource evaluation, we subsample only the training speakers within each training fold while keeping the validation and test folds unchanged. ADReSS and ADReSSo use 20\%, 40\%, and 80\% of the training speakers. Sampling is class-balanced when possible. Speaker subsampling is repeated under different seeds and preserves the no-speaker-overlap constraint. We report mean$\pm$std over seeds after fold-level speaker predictions are computed. For binary ADReSS and ADReSSo classification, we report accuracy, macro-F1, and AUC. For three-class PROCESS classification, we report macro-F1 and Unweighted Average Recall (UAR), which better reflect performance under class imbalance of PROCESS. All metric values are percentages, with the percent sign omitted for brevity.

\noindent\textbf{Implementation details.}
All audio is loaded as mono 16\,kHz waveform and padded within each mini-batch by the wav2vec 2.0 feature extractor. Frozen-probe, merge, and LoRA variants are trained for 30 epochs with the middle-layer readout described above. Full fine-tuning is trained for 15 epochs with a learned weighted sum over layers, giving it a more flexible readout while limiting the severe early overfitting observed on validation folds. All runs use batch size 4, AdamW, weight decay 0.01, and gradient clipping at 1.0. Frozen-probe, merge, and LoRA variants use learning rate $5\times10^{-5}$. Full fine-tuning uses learning rate $1\times10^{-5}$ with 10\% linear warm-up followed by linear decay. LoRA rank is fixed to 8 and is applied to attention and feed-forward linear projections for all LoRA and Dual+LoRA runs. These hyperparameters are shared across datasets and compared methods, except that full fine-tuning unfreezes the encoder while the other variants keep the backbone frozen. We further evaluate layer-pruned dual merging by removing either the first or the last 2/4 transformer layers before training, which probes how sensitive the merged encoder is to different depth regions.

\noindent\textbf{Models:} 
\begin{itemize}[leftmargin=*]
    \item \textbf{Base}: a frozen encoder with a linear probe.
    \item \textbf{Full-FT}: full encoder fine-tuning.
    \item \textbf{Emo} or \textbf{ASR}: single-vector merging of either emotion or ASR.
    \item \textbf{Dual}: the proposed dual merging of both emotion and ASR.
    \item \textbf{Dual+LoRA}: LoRA added on top of dual merging.
    \item \textbf{Layer pruning}: dual-merge variants that remove front or back transformer layers before training.
\end{itemize}

\begin{table}[t]
\centering
\caption{Speaker-independent 5-fold CV on ADReSS and ADReSSo. Results are mean$\pm$std over paired seed-fold runs. Best scores within each dataset are bolded. Parameter counts denote trainable parameters. $\uparrow$: higher the better.}
\label{tab:main}
\setlength{\tabcolsep}{1.5pt}
\renewcommand{\arraystretch}{1.08}
\begin{tabular}{l r ccc}
\toprule
Method & \#Trainable & Acc $\uparrow$ & Macro-F1 $\uparrow$ & AUC $\uparrow$ \\
\midrule
\rowcolor{hlrow}
\multicolumn{5}{l}{\textit{ADReSS}}\\
\rowcolor{white} \textbf{Base (frozen probe)} & 2.1K & 66.2$\pm$2.1 & 65.3$\pm$2.7 & 64.5$\pm$2.8\\
\quad +LoRA                   & 3.5M  & 67.3$\pm$1.1 & 66.7$\pm$1.1 & 63.4$\pm$2.6\\
\quad +Full FT                & 311M  & 62.9$\pm$1.1 & 61.3$\pm$2.3 & 60.5$\pm$1.8\\ \hdashline
\textbf{Merge}               &       &               &               &              \\
\quad +Emo                    & 2.1K  & 67.1$\pm$2.3 & 66.0$\pm$3.4 & 66.3$\pm$1.2\\
\quad +ASR                    & 2.1K  & 69.3$\pm$2.4 & 68.4$\pm$2.6 & 68.5$\pm$1.4\\
\quad +Dual                   & 2.1K  & 69.6$\pm$2.5 & 69.1$\pm$2.8 & \best{69.8$\pm$2.4}\\
\quad +Dual+LoRA              & 3.5M  & \best{71.9$\pm$1.1} & \best{71.5$\pm$1.1} & 69.7$\pm$1.7\\
\midrule
\rowcolor{hlrow}
\multicolumn{5}{l}{\textit{ADReSSo}}\\
\rowcolor{white} \textbf{Base (frozen probe)} & 2.1K & 75.0$\pm$3.7 & 74.3$\pm$3.9 & 79.1$\pm$1.9\\
\quad +LoRA                   & 3.5M  & 79.2$\pm$4.2 & 78.9$\pm$4.3 & 82.3$\pm$4.4\\
\quad +Full FT                & 311M  & 72.0$\pm$6.9 & 70.7$\pm$8.0 & 78.5$\pm$7.9\\ \hdashline
\textbf{Merge}               &       &               &               &              \\
\quad +Emo                    & 2.1K  & 76.4$\pm$4.9 & 75.8$\pm$5.5 & 80.8$\pm$2.4\\
\quad +ASR                    & 2.1K  & 77.1$\pm$5.6 & 76.2$\pm$6.6 & 83.8$\pm$1.7\\
\quad +Dual                   & 2.1K  & 80.8$\pm$4.0 & 80.5$\pm$4.2 & 85.5$\pm$0.7\\
\quad +Dual+LoRA              & 3.5M  & \best{84.3$\pm$4.2} & \best{84.2$\pm$4.3} & \best{87.7$\pm$3.9}\\
\bottomrule
\end{tabular}
\vspace{-5pt}
\end{table}

\subsection{Main Results}

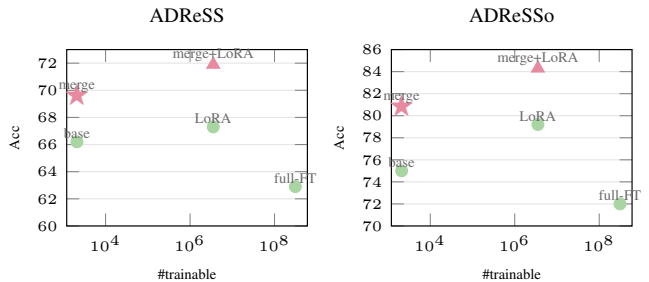
\begin{figure}[t]
\centering
\resizebox{\linewidth}{!}{%
\begin{tabular}{@{}c@{\hspace{2pt}}c@{}}
\begin{tikzpicture}
\begin{axis}[
  width=0.52\linewidth,height=3.7cm,
  title={ADReSS},
  xlabel={\#trainable},ylabel={Acc},
  xmode=log,xmin=1200,xmax=6e8,ymin=60,ymax=73,
  ytick={60,62,64,66,68,70,72},
  ymajorgrids,grid style={black!10},
  tick label style={font=\tiny},label style={font=\tiny},title style={font=\scriptsize},
  nodes near coords,every node near coord/.append style={font=\tiny,color=cGrey,inner sep=1pt},
  point meta=explicit symbolic
]
\addplot[only marks,mark=*,cOther,mark size=2pt] coordinates {(2098,66.2)[base]};
\addplot[only marks,mark=*,cOther,mark size=2pt] coordinates {(3.54e6,67.3)[LoRA]};
\addplot[only marks,mark=*,cOther,mark size=2pt] coordinates {(3.11e8,62.9)[full-FT]};
\addplot[only marks,mark=text,text mark={$\bigstar$},cMergeRed,mark size=3.2pt] coordinates {(2100,69.6)[merge]};
\addplot[only marks,mark=triangle*,cMergeRed,mark size=2.5pt] coordinates {(3.54e6,71.9)[merge+LoRA]};
\end{axis}
\end{tikzpicture}
&
\begin{tikzpicture}
\begin{axis}[
  width=0.52\linewidth,height=3.7cm,
  title={ADReSSo},
  xlabel={\#trainable},ylabel={Acc},
  xmode=log,xmin=1200,xmax=6e8,ymin=70,ymax=86,
  ytick={70,72,74,76,78,80,82,84,86},
  ymajorgrids,grid style={black!10},
  tick label style={font=\tiny},label style={font=\tiny},title style={font=\scriptsize},
  nodes near coords,every node near coord/.append style={font=\tiny,color=cGrey,inner sep=1pt},
  point meta=explicit symbolic
]
\addplot[only marks,mark=*,cOther,mark size=2pt] coordinates {(2098,75.0)[base]};
\addplot[only marks,mark=*,cOther,mark size=2pt] coordinates {(3.54e6,79.2)[LoRA]};
\addplot[only marks,mark=*,cOther,mark size=2pt] coordinates {(3.11e8,72.0)[full-FT]};
\addplot[only marks,mark=text,text mark={$\bigstar$},cMergeRed,mark size=3.2pt] coordinates {(2100,80.8)[merge]};
\addplot[only marks,mark=triangle*,cMergeRed,mark size=2.5pt] coordinates {(3.54e6,84.3)[merge+LoRA]};
\end{axis}
\end{tikzpicture}
\vspace{-5pt}
\end{tabular}%
}
\caption{Accuracy versus trainable parameters. Dual merging occupies a favorable low-capacity region, improving over frozen probe while remaining far smaller than fine-tuning.}
\label{fig:params}
\vspace{-5pt}
\end{figure}

\obs{1) Dual merging improves the frozen acoustic probe.}
Table~\ref{tab:main} shows clear gains from task-vector adaptation. On ADReSS, dual merging raises accuracy from 66.2 to 69.6 (+3.4), macro-F1 from 65.3 to 69.1 (+3.8), and AUC from 64.5 to 69.8 (+5.3). On ADReSSo, accuracy increases from 75.0 to 80.8 (+5.8) and AUC from 79.1 to 85.5 (+6.4). Because the backbone is frozen in both the base and merging systems, these gains come from changing the effective encoder weights through transferred task directions rather than from target-task backbone training. Fig.~\ref{fig:params} further shows that this gain is obtained in a favorable low-capacity region, with roughly the same trainable budget as the frozen probe and far fewer trainable parameters than full fine-tuning. When additional adapter capacity is allowed, Dual+LoRA reaches 71.9 accuracy on ADReSS and 84.3 accuracy on ADReSSo, giving the strongest binary-task results overall. ADReSSo has larger standard deviations because its longer natural recordings yield more variable speaker-level chunk distributions, but under the same folds and seeds, dual merging improves all three metrics over the frozen probe.

\obs{2) Emotion and ASR vectors are complementary.}
Single-vector merging is useful but incomplete. On ADReSS, the emotion vector improves AUC over the base (66.3 vs. 64.5), while the ASR vector gives a larger gain in accuracy (69.3 vs. 66.2). Combining both gives the best balance among merge-only variants and the highest ADReSS AUC. The same trend appears on ADReSSo, where the dual vector reaches 80.8 accuracy and 85.5 AUC. This supports the central design choice. AD speech benefits from both paralinguistic and linguistic knowledge.

\obs{3) Full fine-tuning is not the right default for this data regime.}
Full fine-tuning exposes 311M parameters to each small speaker fold and underperforms the frozen probe on ADReSS (62.9 vs. 66.2 accuracy). By contrast, dual merging trains only 2.1K parameters and still changes the encoder through Eq.~\eqref{eq:merge}. The result is a practical middle ground. It is more adaptive than a probe, but much less prone to target-set overfitting than full fine-tuning.

\obs{4) Merging and LoRA play different roles.}
LoRA alone improves accuracy and macro-F1 over the frozen probe, but it remains below Dual+LoRA on both binary benchmarks. On ADReSS, adding LoRA on top of dual merging gives the strongest accuracy and macro-F1 (71.9/71.5). The same complementarity is clearer on ADReSSo, where Dual+LoRA reaches 84.3 accuracy compared with 79.2 for LoRA and 80.8 for Dual alone. Dual+LoRA is the high-capacity variant of our method, while pure dual merging isolates the effect of reused task knowledge.

\subsection{Further Experiments}

We further examine how dual merging behaves under reduced depth, reduced training data, and the harder PROCESS label space. They test where the same task-vector prior remains useful and where additional capacity is needed.

\begin{table}[t]
\centering
\caption{Layer-pruning analysis on ADReSS. Layers are removed before training, and the classifier reads the midpoint of the remaining hidden-state sequence. $\uparrow$: higher the better. Std omitted for brevity.}
\label{tab:prune}
\setlength{\tabcolsep}{20pt}
\begin{tabular}{l c}
\toprule
Configuration & Acc $\uparrow$\\
\midrule
\textbf{Dual merge}    & 69.6\\ \hdashline
\multicolumn{2}{l}{\textbf{Prune}} \\
\quad -front 2 layers  & 65.1\\
\quad -front 4 layers  & 64.5\\
\quad -back 2 layers   & 69.1\\
\quad -back 4 layers   & 69.1\\
\bottomrule
\end{tabular}
\vspace{-5pt}
\end{table}

\obs{1) Layer pruning shows where transferred cues are used.}
Table~\ref{tab:prune} probes how sensitive dual merging is to removing layers before training. Removing the first 2 or 4 transformer layers reduces ADReSS accuracy to 65.1 and 64.5, respectively, below the unpruned score of 69.6. Removing the last 2 or 4 layers is less harmful, both yielding 69.1 accuracy. Because the classifier always reads the midpoint of the remaining hidden-state sequence, this result should not be read as a causal localization of AD cues in the final layers. It instead shows that the proposed readout relies more on representations formed by the lower-to-middle part of the retained encoder than on the deepest transformer blocks.

\obs{2) Dual merging gives consistent mean gains in low-resource settings.}
Table~\ref{tab:lowres} reports mean accuracy only to keep the low-resource comparison compact. The main full-data tables retain mean$\pm$std. Dual merging outperforms the frozen probe at every budget on ADReSS (63.3 vs. 60.4 with 20\%, 66.3 vs. 64.3 with 40\%, and 67.4 vs. 66.5 with 80\%) and ADReSSo (69.5 vs. 61.0 at 20\%, 73.7 vs. 66.8 at 40\%, and 78.6 vs. 74.6 at 80\%). The point is not that fewer training data make the method better. The result shows that the transferred task-vector prior remains helpful when labeled AD speech is scarce.

\begin{table}[t]
\centering
\caption{Low-resource accuracy on ADReSS and ADReSSo. Values are mean accuracy over seeds. Std omitted for brevity.}
\label{tab:lowres}
\begin{tabular}{ll cccc}
\toprule
Dataset & Method & 20\% & 40\% & 80\% & Full\\
\midrule
\multirow{2}{*}{ADReSS}
& Base & 60.4 & 64.3 & 66.5 & 66.2\\
& Dual & \textbf{63.3} & \textbf{66.3} & \textbf{67.4} & \textbf{69.6}\\
\midrule
\multirow{2}{*}{ADReSSo}
& Base & 61.0 & 66.8 & 74.6 & 75.0\\
& Dual & \textbf{69.5} & \textbf{73.7} & \textbf{78.6} & \textbf{80.8}\\
\bottomrule
\end{tabular}
\end{table}

\begin{table}[t]
\centering
\caption{PROCESS three-class classification (HC/MCI/dementia), speaker-independent cross validation. $\uparrow$: higher the better.}
\label{tab:process}
\setlength{\tabcolsep}{12pt}
\renewcommand{\arraystretch}{1.05}
\begin{tabular}{l cc}
\toprule
Method & Macro-F1 $\uparrow$ & UAR $\uparrow$\\
\midrule
\textbf{Base}    & 42.1$\pm$9.9 & 46.2$\pm$10.9\\
\quad +LoRA      & 46.3$\pm$9.0 & \best{48.8$\pm$8.7}\\
\quad +Full FT   & 39.6$\pm$9.3 & 43.8$\pm$7.4\\ \hdashline
\multicolumn{3}{l}{\textbf{Merge}}\\[-1pt]
\quad +Emo       & 43.0$\pm$11.4 & 46.8$\pm$11.7\\
\quad +ASR       & 42.2$\pm$10.6 & 45.7$\pm$10.7\\
\quad +Dual      & 44.2$\pm$12.3 & 47.0$\pm$11.4\\
\quad +Dual+LoRA & \best{47.4$\pm$10.1} & 48.2$\pm$9.0\\
\bottomrule
\end{tabular}
\vspace{-5pt}
\end{table}

\obs{3) PROCESS favors the combined Dual+LoRA variant.}
Table~\ref{tab:process} reports the three-class PROCESS results. PROCESS is substantially harder than ADReSS-style binary classification because it introduces an intermediate MCI class, severe class imbalance, and finer discrimination along an early cognitive-decline continuum. Pure dual merging improves the frozen probe from 42.1 to 44.2 macro-F1 and from 46.2 to 47.0 UAR, outperforming either single vector. In paired seed-fold comparisons, Dual improves macro-F1 over the base in 12 of 15 runs, while the UAR gain is smaller. The benefit is clearer when the task-vector prior is paired with LoRA. Dual+LoRA gives the best macro-F1 (47.4), slightly above LoRA alone (46.3), and remains close in UAR (48.2 vs. 48.8). The large standard deviations reflect the difficulty of this imbalanced three-way protocol: MCI lies between HC and dementia, and the effective number of speakers per class in each fold is small. As a result, a few speakers near the class boundary can substantially change fold-level macro-F1 and UAR.

\section{Conclusion}
\label{sec:conclusion}

We proposed Dual Task Vector Merging for low-resource AD detection from speech. By injecting emotion and ASR task vectors, the method adapts a frozen wav2vec 2.0 encoder with only layer-wise mixing scalars and a linear head. Across ADReSS and ADReSSo, it improves over frozen probes, LoRA, and full fine-tuning, and remains effective under reduced training data. On PROCESS, Dual+LoRA gives the best macro-F1, showing that task-vector priors can extend beyond binary AD/HC classification when combined with adapter capacity. Task-vector transfer offers a compact and effective route for adapting pretrained speech encoders to low-resource clinical speech tasks.
Future work will validate the method on Chinese AD speech data and further examine its cross-lingual transferability across languages.

\section*{Acknowledgment}
GPT 5.5 was used to assist with language polishing.

\bibliographystyle{IEEEbib}
\bibliography{refs}

@inproceedings{luz2020adress,
  title={Alzheimer's Dementia Recognition through Spontaneous Speech: The {ADReSS} Challenge},
  author={Luz, Saturnino and Haider, Fasih and de la Fuente, Sofia and Fromm, Davida and MacWhinney, Brian},
  booktitle={Proc. Interspeech}, year={2020}
}

@inproceedings{luz2021adresso,
  title={Detecting Cognitive Decline Using Speech Only: The {ADReSSo} Challenge},
  author={Luz, Saturnino and Haider, Fasih and de la Fuente, Sofia and Fromm, Davida and MacWhinney, Brian},
  booktitle={Proc. Interspeech}, year={2021}
}

@article{delafuente2020review,
  title={Artificial Intelligence, Speech, and Language Processing Approaches to Monitoring Alzheimer's Disease: A Systematic Review},
  author={de la Fuente Garcia, Sofia and Ritchie, Craig W. and Luz, Saturnino},
  journal={Journal of Alzheimer's Disease},
  doi={10.3233/JAD-200888},
  year={2020}
}

@inproceedings{cummins2020comparison,
  title={A Comparison of Acoustic and Linguistics Methodologies for Alzheimer's Dementia Recognition},
  author={Cummins, Nicholas and Pan, Yilin and Ren, Zhao and Fritsch, Julian and Nallanthighal, Venkata Srikanth and Christensen, Heidi and Blackburn, Daniel and Schuller, Bj{\"o}rn and Magimai-Doss, Mathew and Strik, Helmer and H{\"a}rm{\"a}, Aki},
  booktitle={Proc. Interspeech},
  doi={10.21437/Interspeech.2020-2635},
  year={2020}
}

@inproceedings{gauder2021speech_embeddings,
  title={Alzheimer Disease Recognition Using Speech-Based Embeddings From Pre-Trained Models},
  author={Gauder, Lara and Pepino, Leonardo and Ferrer, Luciana and Riera, Pablo},
  booktitle={Proc. Interspeech},
  doi={10.21437/Interspeech.2021-753},
  year={2021}
}

@inproceedings{zhu2021wavbert,
  title={{WavBERT}: Exploiting Semantic and Non-Semantic Speech Using Wav2vec and {BERT} for Dementia Detection},
  author={Zhu, Youxiang and Obyat, Abdelrahman and Liang, Xiaohui and Batsis, John A. and Roth, Robert M.},
  booktitle={Proc. Interspeech},
  doi={10.21437/Interspeech.2021-332},
  year={2021}
}

@inproceedings{pappagari2021lowresource,
  title={Automatic Detection and Assessment of Alzheimer Disease Using Speech and Language Technologies in Low-Resource Scenarios},
  author={Pappagari, Raghavendra and Cho, Jaejin and Joshi, Sonal and Moro-Vel{\'a}zquez, Laureano and {\.Z}elasko, Piotr and Villalba, Jes{\'u}s and Dehak, Najim},
  booktitle={Proc. Interspeech},
  doi={10.21437/Interspeech.2021-1850},
  year={2021}
}

@article{baevski2020wav2vec2,
  title={wav2vec 2.0: A Framework for Self-Supervised Learning of Speech Representations},
  author={Baevski, Alexei and Zhou, Yuhao and Mohamed, Abdelrahman and Auli, Michael},
  journal={Advances in Neural Information Processing Systems (NeurIPS)}, year={2020}
}

@inproceedings{liu2024cleverhans,
  title={Clever Hans Effect Found in Automatic Detection of Alzheimer's Disease through Speech},
  author={Liu, Yin-Long and Feng, Rui and Yuan, Jia-Hong and Ling, Zhen-Hua},
  booktitle={Proc. Interspeech},
  doi={10.21437/Interspeech.2024-1018},
  year={2024}
}

@inproceedings{ilharco2023task,
  title={Editing Models with Task Arithmetic},
  author={Ilharco, Gabriel and Ribeiro, Marco Tulio and Wortsman, Mitchell and Schmidt, Ludwig and Hajishirzi, Hannaneh and Farhadi, Ali},
  booktitle={International Conference on Learning Representations (ICLR)}, year={2023}
}

@inproceedings{wortsman2022soups,
  title={Model Soups: Averaging Weights of Multiple Fine-Tuned Models Improves Accuracy without Increasing Inference Time},
  author={Wortsman, Mitchell and Ilharco, Gabriel and Gadre, Samir Ya and others},
  booktitle={International Conference on Machine Learning (ICML)}, year={2022}
}

@inproceedings{yadav2023ties,
  title={{TIES}-Merging: Resolving Interference When Merging Models},
  author={Yadav, Prateek and Tam, Derek and Choshen, Leshem and Raffel, Colin and Bansal, Mohit},
  booktitle={Advances in Neural Information Processing Systems (NeurIPS)}, year={2023}
}

@inproceedings{yang2024adamerging,
  title={{AdaMerging}: Adaptive Model Merging for Multi-Task Learning},
  author={Yang, Enneng and Wang, Zhenyi and Shen, Li and others},
  booktitle={International Conference on Learning Representations (ICLR)}, year={2024}
}

@inproceedings{matena2022fisher,
  title={Merging Models with Fisher-Weighted Averaging},
  author={Matena, Michael and Raffel, Colin},
  booktitle={Advances in Neural Information Processing Systems (NeurIPS)}, year={2022}
}

@article{adaltm2026,
  title={{AdaLTM}: Adaptive Layer-wise Task-Vector Merging of {ASR} Knowledge for Speech Emotion Recognition},
  author={Lee, Chia-Yu and Chou, Huang-Cheng and Lin, Tzu-Quan and Li, Yuanchao and Wu, Ya-Tse and Narayanan, Shrikanth and Lee, Chi-Chun},
  journal={Proc. Interspeech 2026},
  year={2026}
}

@inproceedings{li2025semicog,
  title={Semi-Supervised Cognitive State Classification from Speech with Multi-View Pseudo-Labeling},
  author={Li, Yuanchao and Zhang, Zixing and Han, Jing and Bell, Peter and Lai, Catherine},
  booktitle={IEEE ICASSP},
  doi={10.1109/ICASSP49660.2025.10890660},
  year={2025}
}

@article{liu2026asrerrors,
  title={Beyond Manual Transcripts: Exploring the Potential of Automatic Speech Recognition Errors in Improving {Alzheimer's} Disease Detection},
  author={Liu, Yin-Long and Li, Yuanchao and Feng, Rui and Chen, Jiaxin and Wang, Yiming and Chen, Yu-Ang and Ding, Nan and Yuan, Jiahong and Ling, Zhen-Hua},
  journal={Journal of Biomedical Informatics},
  volume={173},
  doi={10.1016/j.jbi.2025.104968},
  year={2026}
}

@inproceedings{hu2022lora,
  title={{LoRA}: Low-Rank Adaptation of Large Language Models},
  author={Hu, Edward J and Shen, Yelong and Wallis, Phillip and Allen-Zhu, Zeyuan and Li, Yuanzhi and Wang, Shean and Wang, Lu and Chen, Weizhu},
  booktitle={International Conference on Learning Representations (ICLR)}, year={2022}
}

@inproceedings{tao2024process,
  title={Early Dementia Detection Using Multiple Spontaneous Speech Prompts: The {PROCESS} Challenge},
  author={Tao, Fuxiang and Mirheidari, Bahman and Pahar, Madhurananda and Young, Sophie and Xiao, Yao and Elghazaly, Hend and Peters, Fritz and Illingworth, Caitlin and Braun, Dorota and O'Malley, Ronan and Bell, Simon and Blackburn, Daniel and Haider, Fasih and Luz, Saturnino and Christensen, Heidi},
  booktitle={IEEE ICASSP},
  doi={10.1109/ICASSP49660.2025.10889017},
  year={2025}
}

@inproceedings{ramesh2024task,
  title={Task Vector Algebra for {ASR} Models},
  author={Ramesh, Gowtham and Audhkhasi, Kartik and Ramabhadran, Bhuvana},
  booktitle={Proc. IEEE ICASSP},
  doi={10.1109/ICASSP48485.2024.10447848},
  year={2024}
}

@article{lin2026speechft,
  title={{Speech-FT}: Merging Pre-Trained and Fine-Tuned Speech Representation Models for Cross-Task Generalization},
  author={Lin, Tzu-Quan and Huang, Wei-Ping and Tang, Hao and Lee, Hung-yi},
  journal={IEEE Transactions on Audio, Speech and Language Processing},
  doi={10.1109/TASLPRO.2025.3635827},
  year={2026}
}

@inproceedings{hsu2021robust,
  title={Robust wav2vec 2.0: Analyzing Domain Shift in Self-Supervised Pre-Training},
  author={Hsu, Wei-Ning and Sriram, Anuroop and Baevski, Alexei and Likhomanenko, Tatiana and Xu, Qiantong and Pratap, Vineel and Kahn, Jacob and Lee, Ann and Collobert, Ronan and Synnaeve, Gabriel and Auli, Michael},
  booktitle={Proc. Interspeech},
  doi={10.21437/Interspeech.2021-236},
  year={2021}
}

\end{document}